\documentclass[aps,pre,reprint,superscriptaddress]{revtex4-1}

\pdfoutput=1

\usepackage{graphicx}
\usepackage{amssymb,amsmath}
\usepackage{bm}

\begin{document}

\title{Mechanical relaxation and fracture of phase field crystals}
\author{Wenquan Zhou}
\affiliation{State Key Laboratory of Solidification Processing,
  Northwestern Polytechnical University, Xi'an 710072, China}
\author{Jincheng Wang}
\email{jchwang@nwpu.edu.cn}
\affiliation{State Key Laboratory of Solidification Processing,
  Northwestern Polytechnical University, Xi'an 710072, China}
\author{Zhijun Wang}
\affiliation{State Key Laboratory of Solidification Processing,
  Northwestern Polytechnical University, Xi'an 710072, China}
\author{Zhi-Feng Huang}
\email{huang@wayne.edu}
\affiliation{Department of Physics and Astronomy, Wayne State University,
  Detroit, Michigan 48201, USA}

\date{\today}

\begin{abstract}
  A computational method is developed for the study of mechanical response and
  fracture behavior of phase field crystals (PFC), to overcome a limitation of
  the PFC dynamics which lacks an effective mechanism for describing fast mechanical
  relaxation of the material system. The method is based on a simple interpolation
  scheme for PFC (IPFC) making use of a condition of the displacement field to satisfy
  local elastic equilibration, while preserving key characteristics of the original
  PFC model. We conduct a systematic study on the mechanical properties of a sample
  nanoribbon system with honeycomb lattice symmetry subjected to uniaxial tension,
  for numerical validation of the IPFC scheme and the comparison with the original
  PFC and modified PFC methods. Results of mechanical response, in both elasticity
  and fracture regimes, show the advantage and efficiency of the IPFC method across
  different system sizes and applied strain rates, due to its effective process
  of mechanical equilibration. A brittle fracture behavior is obtained in IPFC
  calculations, where effects of system temperature and chirality on the fracture
  strength and Young's modulus are also identified, with results agreeing with
  those found in previous atomistic simulations of graphene. The IPFC scheme
  developed here is generic and applicable to the mechanical studies using different
  types of PFC free energy functionals designed for various material systems.
\end{abstract}

\maketitle

\section{Introduction}

The mechanical properties of materials, including their strength, fracture behavior,
and elastic properties, are among the key factors determining the technological
applications of the material system. A recent example is the exploration of novel
two-dimensional (2D) monolayer materials such as graphene, which is known as one of
the strongest materials being made. Both experimental \cite{lee08} and computational
\cite{ZhaoNanoLett09,ZhaoJAP10,grantab10,liu07,ZakharchenkoPRL09}
efforts have been devoted to the study of its mechanical and fracture
behaviors under various sample conditions such as system temperature, chirality,
defects, sample size, imposed strain rate, and crack length.

Computational studies of these material mechanical behaviors are often relied
upon the use of atomistic simulation techniques, including molecular dynamics (MD)
\cite{ZhaoNanoLett09,ZhaoJAP10,grantab10}, first-principles density functional
theory (DFT) \cite{liu07}, and Monte Carlo (MC) \cite{ZakharchenkoPRL09} methods.
These atomistic methods are usually characterized by the microscopic length and
atomic vibration time scales, and thus are limited by the small system sizes and
short dynamic time ranges that they can access. Although such constraints on spatial
and temporal scales can be significantly relieved by the use of continuum approaches,
the traditional continuum theories, such as phase field models and continuum elasticity
theory, lack the explicit details of atomic-level crystalline microstructures
that are important for determining the mechanical and fracture properties of material
systems particularly those beyond the elasticity regime.

To overcome this difficulty in materials modeling, one of recent efforts has been
put on combining the microscopic crystalline details with continuum density-field
formulation, in particular the development of phase field crystal (PFC) models
\cite{PhysRevLett.88.245701,PhysRevE.70.051605,PhysRevLett.96.225504,PhysRevLett.105.045702,
Mkhonta13,Wang18r} which have attracted large amounts of recent research interest
and resulted in a wide range of applications such as the study of solidification
\cite{PhysRevE.70.051605,PhysRevLett.105.045702,Mkhonta13}, defect structures
and dynamics \cite{PhysRevE.73.031609,PhysRevLett.118.255501,PhysRevB.87.024110,
SalvalaglioPRMater18,SkaugenPRB18}, crystal nucleation and grain growth
\cite{Wu12,0953-8984-22-36-364101,GUO2018175}, surface ordering and patterns
\cite{Elder16,Smirman17}, mechanical behavior or crack dynamics
\cite{PhysRevLett.96.225504,PhysRevE.80.046107,ZHOU2017121,0965-0393-24-5-055010,
Hu2017}, among many others. Although the PFC method has the advantage of being able to
address large length and time scales without losing atomic spatial resolution of the
microstructure, the standard dynamics of PFC is characterized only by slow, diffusive
time scales for all the system dynamical processes (including the elastic relaxation).
It thus lacks a separate mechanism for elastic and mechanical response, for
which the relaxation process occurs on much smaller phonon-level time scales or
almost instantaneously (i.e., with instantaneous mechanical equilibrium), a
feature that is particularly important for the study of material mechanical
behavior and fracture.

This drawback of PFC dynamics can be partially remedied by incorporating the
damped wave modes into the PFC equation, i.e., the modified PFC (MPFC) model
\cite{PhysRevLett.96.225504,PhysRevE.80.046107} which, however, is still limited
by the associated effective length scale of elastic interaction and by its stricter
requirement of numerical convergence in computation. The model is also subjected to
a restriction on the fastness of system dynamics it can achieve \cite{HeinonenPRE16}.
In principle, the dynamics of mechanical relaxation should be examined through the
details of acoustic phonon modes. The incorporation of them in the PFC framework
has been derived via the Poisson bracket formalism \cite{MajaniemiPRB07}, and
usually manifests in terms of hydrodynamic couplings (see, e.g., the hydrodynamic
models \cite{BaskaranCMAME16,TothJPCM14} based on classical DFT of freezing and
the Navier-Stokes equation), although further work is needed to identify the effect
and computational efficiency of these models on the mechanical relaxation processes.
Recently much attention has been paid to the construction of amplitude equation
formulation for PFC, with two methods developed to address the issue of fast elastic
relaxation \cite{HeinonenPRL16,HeinonenPRE16,Heinonen14}. The first one is through
imposing a separate, extra condition of elastic equilibration (via the phases of
complex amplitudes) on the standard overdamped, dissipative dynamics governing the
slowly varying amplitudes of the density field \cite{Heinonen14}. In the second
method hydrodynamic coupling is introduced to the PFC amplitude formulation,
such that the elastic relaxation is determined by large-wavelength phonon modes
\cite{HeinonenPRL16,HeinonenPRE16}. Both methods can well produce the fast dynamics
for mechanical equilibration, but the corresponding slow-scale amplitude description
does not incorporate the coupling to the underlying microscopic lattice structure and
thus lacks the resulting lattice pinning effect \cite{Huang13} and Peierls barriers
for defect motion \cite{PhysRevE.73.031609}, which are needed for understanding the
mechanical behavior of materials. In addition, a very recent study showed that an
additional smooth distortion field and the associated compatible strain should be
incorporated into PFC to obtain full mechanical equilibrium of the system
\cite{SkaugenArXiv18}. The method is based on linear elasticity and is to be extended
to the nonlinear elasticity regime. In short, so far a complete PFC-type dynamics that
can cover the full range of system characteristic time and length scales and the real
material evolution processes (ranging from fast elastic to slow diffusive relaxation)
is still lacking, and the solution is elusive given the coarse-grained nature (in both
space and time) of the PFC-type density field approach.

In this work we introduce an alternative, effective computational scheme for PFC mechanical
relaxation, via imposing an additional constraint on the original PFC model by assuming
that after each step of mechanical deformation, the system would instantaneously adjust
to a state close to local elastic equilibrium and then relax from this new initial state
to reach the mechanical equilibration. It is achieved by a simple interpolation algorithm
for the PFC density field, based on the property of linear spatial dependence of atomic
displacements under small strain increment in between two subsequent deformation steps.
This facilitates a rapid relaxation to the mechanical equilibrium state of the system
even with the use of standard diffusive PFC dynamics. The validity and high efficiency
of this method are verified through the study of uniaxial tensile test on a sample
double-notched nanoribbon system with 2D honeycomb lattice symmetry. Effects of system
size, strain rate, temperature, and structure chirality have been systematically examined,
with results of mechanical response and fracture compared to the calculations using the
original PFC and MPFC models, to demonstrate the advantage of this interpolation scheme
of PFC (IPFC) particularly for large system sizes and a broad range of strain rates.

\section{Models and Method}

\subsection{PFC and MPFC models}

In the original PFC model for single-component systems \cite{PhysRevLett.88.245701,PhysRevE.70.051605},
the dimensionless free energy functional is given by
\begin{equation}\label{1101}
  F=\int d\vec{r}\left\{\frac{\phi}{2}[r+(\nabla+1)^2]\phi
  +\frac{\tau}{3}\phi^3+\frac{1}{4}\phi^4\right\},
\end{equation}
where the order parameter $\phi$ represents the variation of the atomic number density field from
a constant reference value, and $r$ and $\tau$ are phenomenological parameters. The equilibrium
thermodynamic properties of the PFC model can be controlled by varying the temperature parameter
$r$ and the average atomic density \cite{PhysRevE.70.051605},
showing a transition between the homogeneous (or liquid) phase to the spatially periodic,
crystalline solid state. In a crystalline solid phase, the order parameter field $\phi$ can
be written in a general form:
\begin{equation}\label{1102}
 \phi(\vec r)=\phi_{0}+\sum_{n,m} A_{nm} e^{i\vec G_{nm} \cdot \vec r},
\end{equation}
where $\phi_{0}$ is the average atomic density variation, $A_{nm}$ are the amplitudes,
and $\vec G_{nm} = n\vec b_{1}+m\vec b_{2}$, with $(n,m)$ the Miller indices
and $(\vec b_{1}, \vec b_{2})$ the principle reciprocal lattice vectors. For the example of
a 2D lattice with hexagonal symmetry, we have
\begin{equation}\label{1103}
  \vec b_{1} = q_0 \left ( \frac{\sqrt{3}}{2} \hat x + \frac{1}{2}\hat y \right ),
  \qquad \vec b_{2} = q_0 \hat y,
\end{equation}
where $q_0={2\pi}/(a_{0}\sqrt{3}/2)$ with the lattice constant $a_{0}$. Within the one-mode
approximation $(n,m)=(\pm 1, 0)$, $(0, \pm 1)$, and $(\pm 1, \mp 1)$, $A_{\pm 1 0} = A_{\pm 1 \mp 1}
\equiv A_{\rm eq}$, $A_{0 \pm 1} = -A_{\rm eq}$, and thus
\begin{equation}\label{1104}
  \phi=\phi_{0}+2A_{\rm eq} \left [2 \cos \left (\frac{\sqrt{3}}{2} q_0 x \right )
    \cos \left ( \frac{1}{2}q_0y \right ) - \cos(q_0 y) \right ].
\end{equation}
In the equilibrium state determined by the free energy minimization, $q_0=1$, and
when $\tau+3\phi_{0}>0$, 
\begin{equation}\label{1105}
A_{\rm eq}=-\frac{1}{15} \left ( \tau+3\phi_{0}+\sqrt{\tau^2-15r-24\tau\phi_{0}-36\phi_{0}^2} \right ),
\end{equation}
corresponding to a 2D honeycomb lattice as will be examined below. Note that it is essentially
the inverse of triangular phase in one-mode PFC \cite{Elder16}.

The standard PFC dynamics is of dissipative nature and governed by the conserved, time-dependent
Ginzburg-Landau equation $\partial \phi / \partial t = \nabla^2 \delta F / \delta \phi$,
which leads to
\begin{equation}\label{1106}
 \frac{\partial\phi}{\partial t}=\nabla^2 \left [r\phi+(\nabla^2+1)^2\phi+\tau\phi^2+\phi^3 \right ].
\end{equation}
The corresponding system evolution are then controlled by diffusive dynamics, even for
processes that are related to much faster time scales such as elastic or plastic relaxation
and mechanical response. To overcome this shortcoming, the above PFC dynamics has been modified
by adding a wave-mode term of second order time derivative \cite{PhysRevLett.96.225504},
so that two different time scales of the diffusional and elastic or phonon-type modes can
be incorporated, i.e.,
\begin{equation}\label{1121}
  \frac{\partial^2\phi}{\partial t^2}+ \beta \frac{\partial\phi}{\partial t}
  =\alpha^2 \nabla^2 \left [r\phi+(\nabla^2+1)^2\phi+\tau\phi^2+\phi^3 \right ],
\end{equation}
where $\alpha$ is proportional to the speed of sound wave and $\beta$ is a phenomenological
parameter associated with the damping rate. It is important to know that in this MPFC model,
the fast elastic propagating behavior is limited within an effective elastic interaction
range that is proportional to $\alpha/\beta$ \cite{PhysRevLett.96.225504,PhysRevE.80.046107},
beyond which the diffusive dynamics (similar to that of original PFC) dominates. Also,
recent analysis indicated that although the elastic interaction range can always be
increased by reducing the $\beta$ value, beneath a certain threshold of $\beta$ the
system dynamics cannot be further accelerated \cite{HeinonenPRE16}. Such limitations
play an important role on the size effect and computational efficiency for the simulation
of system mechanical response, as will be demonstrated below.

\subsection{Modeling of systems under uniaxial tension}
\label{sec:setup}

To model a PFC or MPFC system subjected to a uniaxial tension, we use the traction boundary
conditions introduced in Refs. \cite{PhysRevLett.96.225504,PhysRevE.80.046107} by adding an
additional energy penalty term $F_{\rm ext}$ into the free energy functional Eq.~(\ref{1101}),
which effectively fixes the density field of the loaded and moving boundary layers (the
traction regions) to a pre-defined field $\phi_{\rm trac}$ representing the corresponding
equilibrium crystal structure, i.e.,
\begin{equation}\label{2092}
  F_{\rm ext}=\int d{\vec r} M(\vec r) \left [ \phi(\vec r)-\phi_{\rm trac}(\vec r) \right ]^2.
\end{equation}
Here the traction function $M(\vec r)$ is set to be a positive constant ($=2$) within the
traction regions that are moved with a specific strain rate $\dot{\varepsilon}$, and
is zero outside the regions. In our simulations $\phi_{\rm trac}$ is chosen as the equilibrium
profile of the traction regions determined numerically before the imposing of tension.

\begin{figure}
\includegraphics[width=0.3\textwidth]{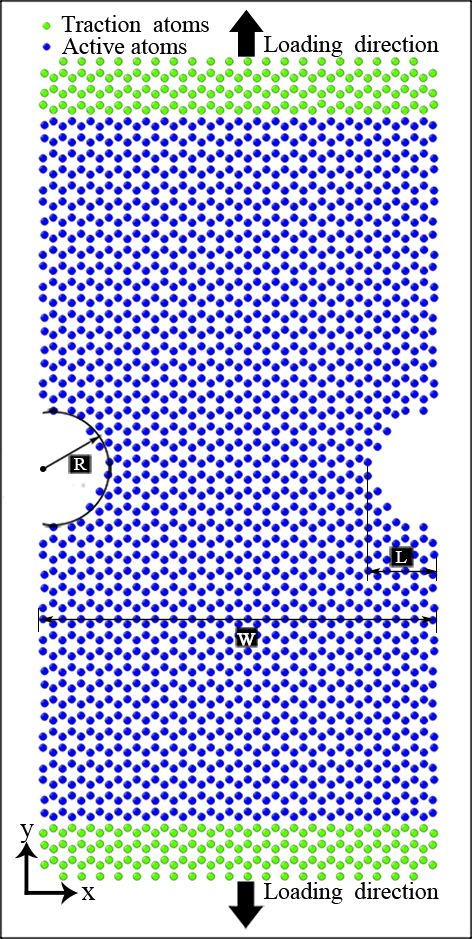}
\caption{Schematic of a 2D double notched single-crystal nanoribbon subjected to a
  uniaxial tensile loading.}
\label{fig:ff1}
\end{figure}

A schematic of the corresponding system setup for mechanical deformation is given in
Fig.~\ref{fig:ff1}. What is illustrated there is an example system consisting of two
types of crystalline regions (blue and green) surrounded by a coexisting homogeneous
phase (white margins). The solid sample is stretched vertically along the $y$ direction
at both ends, with the tensile load applied on seven rows of atoms at each end (green;
traction regions with $M(\vec r) > 0$). The mechanical relaxation occurs in the middle
solid region (blue; active zone with $M(\vec r)=0$), which is configured as a double
notched single-crystal nanoribbon for our tests. Before the application of the uniaxial
tension, the whole system is relaxed to reach an equilibrium state which is used as the
initial condition of the subsequent tensile test, with the initial lengths of the active
zone denoted as $L_x^0$ and $L_y^0$ in the $x$ and $y$ directions, respectively. For the
example of Fig.~\ref{fig:ff1}, the total system size is $256 \Delta x \times 512 \Delta y$,
while the initial active region of the nanoribbon is measured as $L_x^0 \times L_y^0 =
196\Delta x \times 340\Delta y$. Various system sizes have been used in our simulations,
all with similar system setup.

To identify the mechanical properties during the tensile test, we calculate the strain
energy of the nanoribbon as $F_s=F(\varepsilon_e)-F(\varepsilon_e=0)$, where
$\varepsilon_e=(L_y-L_y^0)/L_y^0$ is the applied engineering strain, with $L_y$ the
strained length of the active zone along the stretching direction. The engineering
stress is calculated by
\begin{equation}
  \sigma_e = \frac{1}{\mathcal{A}_0} \frac{\partial F_s}{\partial \varepsilon_e}
  = \frac{1}{\dot{\varepsilon}} \frac{\partial (F_s/\mathcal{A}_0)}{\partial t},
  \label{eq:sigma}
\end{equation}
where the initial area $\mathcal{A}_0 = L_x^0 \times L_y^0$, $F_s/\mathcal{A}_0$ is
the strain energy density, and $\dot{\varepsilon}$ is the strain rate. During each
tensile loading step, the nanoribbon is stretched by a minimum increment of one grid
spacing $\Delta y$ at each end, resulting in a strain increment of $\Delta \varepsilon_e
= 2\Delta y /L_y^0$. If this stretching increment is imposed every $N$ time steps
(i.e., every $N \Delta t$ of time), we have
\begin{equation}
  \dot{\varepsilon} = \frac{\Delta \varepsilon_e}{N \Delta t}
  = \frac{2\Delta y}{L_y^0 N \Delta t}. \label{eq:rate_PFC}
\end{equation}

An extra attention needs to be paid to the time scale and hence the strain rate for
the wave-mode MPFC model as compared to the diffusive PFC dynamics. In the original
PFC model Eq.~(\ref{1106}), the associated time scale can be determined via that of
vacancy diffusion, $\tau_D=a_0^2/D$ \cite{PhysRevE.73.031609}, where $a_0$ is the
lattice spacing [$a_0 = 4\pi/(\sqrt{3}q_0)$ in PFC] and $D$ is the vacancy diffusion
constant determined by $D = 1 + r + 2\tau \phi_0 + 3\phi_0^2 + 18 A_{\rm eq}^2$
\cite{PhysRevE.70.051605}. If labeling all the corresponding dimensional variables
or parameters by a superscript ``$d$'' to distinguish from the dimensionless
quantities in the PFC equations, we can identify the PFC time scale as
\begin{equation}
  \left. \frac{t^d}{t} \right |_{\rm PFC} = \frac{\tau_D^d}{\tau_D}
  = \left ( \frac{a_0^d}{a_0} \right )^2 \left ( \frac{D}{D^d} \right ),
\end{equation}
where $a_0^d$ and $D^d$ are the lattice spacing and vacancy diffusion constant of
the specific real material to be studied. On the other hand, in MPFC governed by
the wave dynamic Eq.~(\ref{1121}), the vacancy diffusion is characterized by an
effective diffusion coefficient $D_{\rm MPFC} = \alpha^2 (1 + r + 2\tau \phi_0
+ 3\phi_0^2 + 18 A_{\rm eq}^2)/\beta = (\alpha^2/\beta) D$ \cite{PhysRevLett.96.225504},
leading to a MPFC time scale
\begin{equation}
  \left. \frac{t^d}{t} \right |_{\rm MPFC}
  = \left ( \frac{a_0^d}{a_0} \right )^2 \left ( \frac{D_{\rm MPFC}}{D^d} \right )
  = \frac{\alpha^2}{\beta} \left. \frac{t^d}{t} \right |_{\rm PFC},
  \label{eq:t_scale}
\end{equation}
an increase by a factor of $\alpha^2/\beta$ compared to the PFC time scale.
Thus this factor needs to be incorporated into the strain rate calculation in
MPFC, i.e.,
\begin{equation}
  \dot{\varepsilon}_{\rm MPFC} = \frac{2 \beta \Delta y}{\alpha^2 L_y^0 N \Delta t},
  \label{eq:rate_MPFC}
\end{equation}
to be comparable with the PFC strain rate Eq.~(\ref{eq:rate_PFC}).

\subsection{An interpolation scheme for mechanical relaxation}

\begin{figure}
\includegraphics[width=0.48\textwidth]{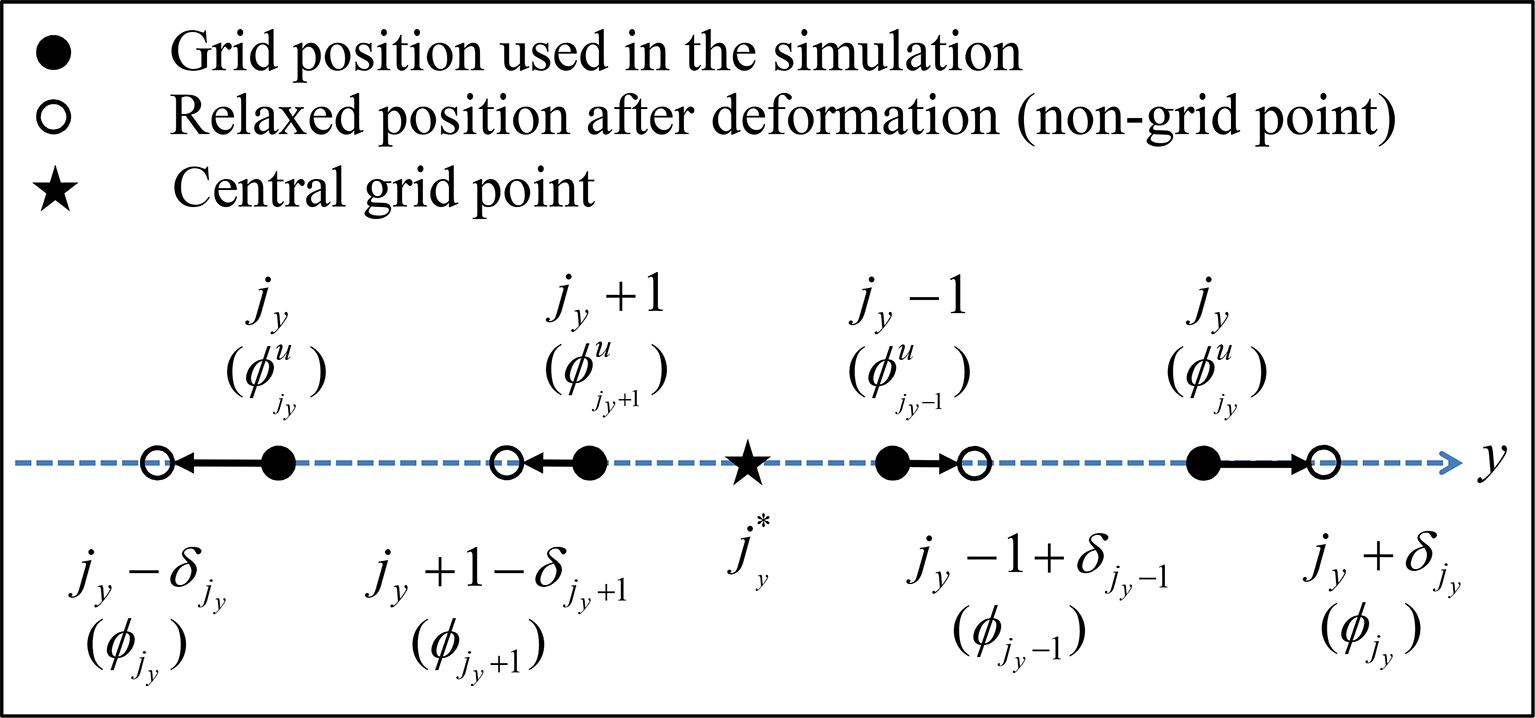}
\caption{Schematic of mechanical deformation along the $y$ direction parallel to the
  uniaxial tensile loading. The star point indicates the grid position at the central
  line of the stretched nanoribbon. Note that the filled points shown here are the
  simulation grid points, but not necessarily the atomic lattice sites, while the open
  points indicate the corresponding relaxed positions after each step of deformation
  which, however, are not grid positions due to $0 < \delta_{j_y} < 1$.}
\label{fig:ff2}
\end{figure}

As discussed above, a key factor for effectively modeling the mechanical response of
a system is to achieve fast (or close to instantaneous) elastic relaxation in system
dynamics, which is the motivation behind the recent development of MPFC
\cite{PhysRevLett.96.225504,PhysRevE.80.046107} and amplitude
\cite{Heinonen14,HeinonenPRL16,HeinonenPRE16}
methods. Here we introduce a simple, alternative algorithm to efficiently facilitate the
rapid process of strain relaxation in systems under mechanical deformation. It makes use of
a property of linear elasticity assuming the linear spatial dependence of the displacement
field $u_y$ when reaching mechanical equilibrium, given a small strain increment
$\Delta \varepsilon$ imposed by each stretching step, i.e.,
\begin{equation}
  u_y \cong \Delta \varepsilon (y-y^*), \qquad \Delta \varepsilon = 2\Delta y / L_y,
  \label{eq:uy}
\end{equation}
where the displacement $u_y$ is measured with respect to the positions at the beginning
of each stretching step, so that after this stretching $y \rightarrow y+u_y$ and
the strained length $L_y$ is increased by $2\Delta y$ given the two-end pulling.
$\Delta \varepsilon$ represents the incremental strain in between two subsequent
steps as a result of the new tensile loading, and $y^*$ is the central position of
the stretched sample with zero displacement. [A similar setup can be used in the
case of one-end pulling for which the fixed end is located at $y=y^*=0$ and thus
$\Delta \varepsilon = \Delta y / L_y$ in Eq.~(\ref{eq:uy}).]
Note that $\Delta \varepsilon$ is always
very small since it accounts for the applied strain increase due to the stretching by
only one grid spacing at each end, and thus Eq.~(\ref{eq:uy}) is a reasonably good
approximation even beyond the stage of elasticity or close to the fracture regime.
A corresponding one-dimensional (1D) schematic along the stretching $y$ direction
is shown in Fig.~\ref{fig:ff2}, indicating (i) the fixed central grid point
$j_y^* = y^*/\Delta y$, (ii) the grid positions $j_y=y/\Delta y$ before the current-step
deformation (filled points), and (iii) the relaxed ones $j_y + \delta_{j_y}$ for $y > y^*$
and $j_y > j_y^*$, or $j_y - \delta_{j_y}$ for $y < y^*$ and $j_y < j_y^*$ after deformation
(open points). Given Eq.~(\ref{eq:uy}), we have
\begin{equation}
  \delta_{j_y} = |u_y|/\Delta y = \Delta \varepsilon |j_y - j_y^*|,
  \label{eq:jy}
\end{equation}
and $0 < \delta_{j_y} < \Delta \varepsilon (L_y/2)/\Delta y = 1$.

Between two consecutive steps of tensile loading, assume $\phi_{j_y}$ is the ``old'' density
field obtained from mechanical relaxation following the previous step (i.e., right before
the new stretching), and $\phi^u_{j_y}$ is the updated density field at the same simulation
grid position $j_y$ after the new tensile load. As a result of the strain-induced linear
displacement described by Eqs.~(\ref{eq:uy}) and (\ref{eq:jy}) with $y \rightarrow y+u_y$,
we get $\phi(y) \rightarrow \phi^u(y+u_y)$, i.e., $\phi^u_{j \pm \delta_{j_y}} = \phi_{j_y}$
(``$+$'': for $j_y > j_y^*$; ``$-$'': for $j_y < j_y^*$). Note that $j \pm \delta_{j_y}$
is not a grid position used in numerical simulation due to $0 < \delta_{j_y} < 1$. We then
determine the value of $\phi^u_{j_y}$ at any grid point $j_y$ by a linear interpolation
either between $j_y-1+\delta_{j_y-1}$ and $j_y+\delta_{j_y}$ when $j_y > j_y^*$, or between
$j_y-\delta_{j_y}$ and $j_y+1-\delta_{j_y+1}$ when $j_y < j_y^*$ (see Fig.~\ref{fig:ff2}).
Therefore, for $j_y < j_y^*$,
\begin{eqnarray}
  \phi_{j_y}^u &=& \phi^u_{j_y-\delta_{j_y}} + \frac{\delta_{j_y}}{1+\delta_{j_y}-\delta_{j_y+1}}
  (\phi^u_{j_y+1-\delta_{j_y+1}} - \phi^u_{j_y-\delta_{j_y}}) \nonumber\\
  &=& \phi_{j_y} + \frac{\delta_{j_y}}{1+\delta_{j_y}-\delta_{j_y+1}}
  \left (\phi_{j_y+1} - \phi_{j_y} \right ), \label{2094}
\end{eqnarray}
while for $j_y >  j_y^*$,
\begin{eqnarray}
  \phi_{j_y}^u &=& \phi^u_{j_y-1+\delta_{j_y-1}} \nonumber\\
  && + \frac{1-\delta_{j_y-1}}{1+\delta_{j_y}-\delta_{j_y-1}}
  \left (\phi^u_{j_y+\delta_{j_y}} - \phi^u_{j_y-1+\delta_{j_y-1}} \right ) \nonumber\\
  &=& \phi_{j_y-1} + \frac{1-\delta_{j_y-1}}{1+\delta_{j_y}-\delta_{j_y-1}}
  \left (\phi_{j_y} - \phi_{j_y-1} \right ). \label{2095}
\end{eqnarray}
In addition, for the conserved dynamics of $\phi$ field the average density of the
overall system is kept unchanged (or equivalently, its zero-mode Fourier component
at $\vec q =0$ be fixed). After then, the system is relaxed and equilibrated
for $N$ time steps via the standard PFC dynamics of Eq.~(\ref{1106}) before
the next tensile loading, with $N$ determined by the PFC strain rate through
Eq.~(\ref{eq:rate_PFC}). As will be demonstrated below from numerical simulations,
this interpolated PFC (IPFC) scheme has the advantage of a much faster elastic
relaxation particularly for large sample size of deformation.

\section{Results and Discussion}

We have conducted a systematic study on the mechanical response of single-crystal
nanoribbons with honeycomb lattice symmetry using three methods of original
PFC model, MPFC, and the IPFC scheme. The simulations are based on the 2D system
setup given in Sec.~\ref{sec:setup} and Fig.~\ref{fig:ff1} for various choices of
system size and strain rate, with periodic boundary conditions applied in both
directions. A pseudospectral algorithm with an exponential propagation scheme
and a predictor-corrector method \cite{Cross94} are used to numerically solve
the PFC equation (\ref{1106}), while for MPFC a similar numerical algorithm is
adopted, with details presented in the Appendix.

For each tensile test, we use the same initial condition for all three different
methods, as prepared by equilibrating the nanoribbon configuration through
Eq.~(\ref{1106}) with standard PFC dynamics (up to $t = 5 \times 10^5 \Delta t$
without external stress). The model parameters are chosen such that the 2D solid
sheet is characterized by sharp and faceted surfaces, and the average densities
in the solid and homogeneous regions are set as the coexisting values determined
from the phase diagram, e.g., $\phi_0=0.1027$ (solid) and $0.3617$ (homogeneous)
for $r=-0.5$ and $\tau=1$ used in Sec.~\ref{sec:comparison}. Such a state of
solid-homogeneous phase coexistence is well maintained during the subsequent
process of tensile loading and fracture, with no extra setting needed. No
additional solidification or melting at the interfaces (including the notches)
and no re-crystallization of the fracture line after it occurs are found in our
simulations, for which the condition of sharp and faceted solid surface
plays an important role. 
Other parameters are set as $\Delta x=\Delta y=\pi/4$ for grid spacing,
$\Delta t=0.4$ (for PFC and IPFC) or $0.001$ (for MPFC), and $(\alpha, \beta)
= (15, 0.9)$ used in MPFC. The spatial resolution of the numerical grid is
kept unchanged throughout the simulation of tensile deformation. In addition,
we have tested other constant values of grid spacing (via e.g., spot checks of
systems with $r=-0.5$ and $\Delta x=\Delta y$ ranging from $\pi/4$ to $\pi/8$),
and obtained very similar results of faceted surfaces and mechanical behavior
(including the stress-strain relation and fracture).

\subsection{Comparison between PFC, MPFC, and IPFC methods}
\label{sec:comparison}

\subsubsection{Effects of system size and strain rate}

\begin{figure*}
\includegraphics[width=\textwidth]{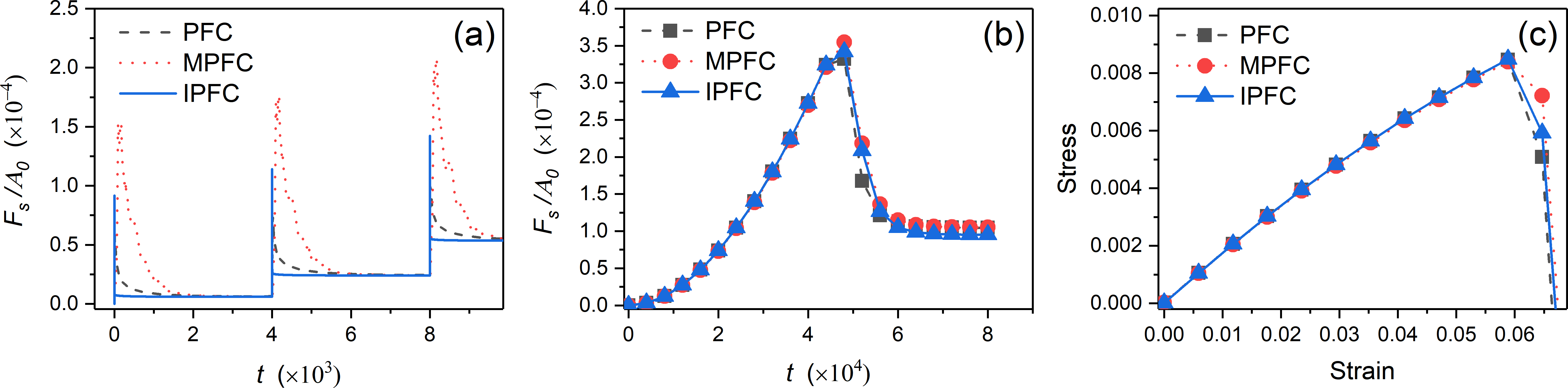}
\caption{Mechanical behavior of the double notched nanoribbon calculated via three different
  methods, for a system grid size $256 \times 512$ (with $\mathcal{A}_0 = L_x^0 \times L_y^0
  = 196 \Delta x \times 340 \Delta y$) and strain rate $\dot{\varepsilon} = 1.471 \times 10^{-6}$.
  (a) Time evolution of strain energy density $F_s/\mathcal{A}_0$ during the first three steps
  of stretching, where the end state of each stretching step after mechanical relaxation is
  used to evaluate the mechanical property of the system in (b) and (c).
  (b) Strain energy density $F_s/\mathcal{A}_0$ evaluated at the mechanically relaxed
  state as a function of time. (c) Stress-strain relation. In (a) and (b), a factor of
  $\alpha^2/\beta$ has been scaled for time $t$ in MPFC results.}
\label{fig:ff3}
\end{figure*}

Figure \ref{fig:ff3} shows the mechanical property of the double notched nanoribbon
(as illustrated in Fig. \ref{fig:ff1}) obtained from PFC, MPFC, and IPFC
simulations. For the small system size $256 \Delta x \times 512 \Delta y$
presented here (particularly the short initial length $L_y^0$ used for stretching),
all three methods yield similar mechanical behavior at small or moderate strain
rates (e.g., $\dot{\varepsilon} = 1.471 \times 10^{-6}$ used in Fig.~\ref{fig:ff3}),
although with different details of strain relaxation. As shown in Fig.~\ref{fig:ff3}(a),
during each tensile loading step, right after the nanoribbon is stretched a sharp peak
appears in the time evolution of strain energy density, which then decreases with
time towards a mechanical equilibrium state.
Among these three methods, the IPFC scheme is most efficient in terms of mechanical
relaxation, with shortest (almost ``instantaneous'') relaxation time to reach the
equilibrium state. This can be attributed to the simple fact that the linear displacement
approximation [Eq.~(\ref{eq:uy})] has been pre-determined in the IPFC scheme at each
stretching step. Although in Fig.~\ref{fig:ff3}(a) the result of standard diffusive PFC
dynamics seems to show a faster elastic relaxation process as compared to wave-mode MPFC,
it should be cautioned that the MPFC result is plotted against the time rescaled by
a factor of $\alpha^2/\beta$ in the figure so that the PFC and MPFC time scales
are matched [see Eq.~(\ref{eq:t_scale})].
Without this rescaling the MPFC relaxation appears much faster than PFC.

The time evolution of mechanically relaxed strain energy density are given
in Fig.~\ref{fig:ff3}(b), which is used to calculate the stress-strain curves in
Fig.~\ref{fig:ff3}(c) based on Eq.~(\ref{eq:sigma}). Very similar stress-strain
relation is obtained for all three methods, particularly in the elastic regime,
although there are some small differences around the fracture stage. A behavior
of brittle fracture is observed in our simulations [see Fig.~\ref{fig:ff3}(c)],
which is qualitatively similar to the MD simulation results for pristine
\cite{ZhaoNanoLett09,ZhaoJAP10} or grain boundary \cite{grantab10} samples
of graphene.

\begin{figure}
\includegraphics[width=0.48\textwidth]{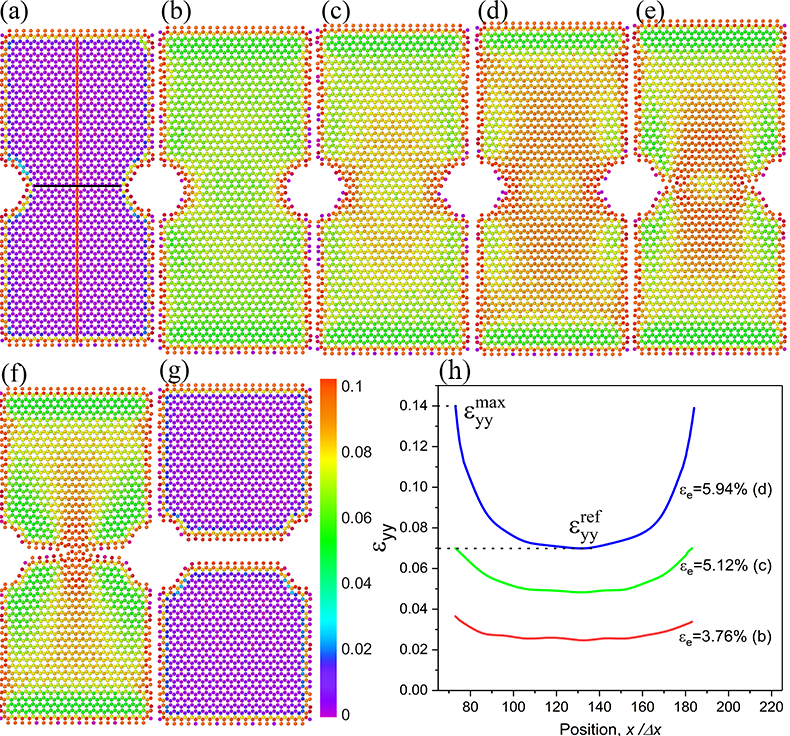}
\caption{Spatial distribution of strain $\varepsilon_{yy}$ over the double notched nanoribbon
  simulated in Fig.~\ref{fig:ff3} using the IPFC scheme, at $\varepsilon_e = 0$, $3.76\%$,
  $5.12\%$, $5.94\%$, $6.82\%$, $7.18\%$, and $9.24\%$ from (a) to (g). (h) Plots of
  $\epsilon_{yy}$ along the central horizontal line of the nanoribbon [black line in (a)]
  at different strain levels.}
\label{fig:ff4}
\end{figure}

The spatial distribution of strain $\varepsilon_{yy}=\partial u_y / \partial y$ in the
double notched sample is presented in Fig.~\ref{fig:ff4}, with results obtained from
IPFC simulation at different stages of imposed tension. A numerical image processing
technique, the Peak Pairs algorithm \cite{GALINDO20071186,WangUltramicroscopy15},
is used to calculate the local strain $\varepsilon_{yy}$. Before the fracture occurs,
the strain is concentrated around the notch roots, as seen in
Figs.~\ref{fig:ff4}(a)--\ref{fig:ff4}(d). This can be quantified by the
stress-concentration factor $K_t$ measuring the ratio between the maximum stress
$\sigma^{\rm max}_{yy}$ at the notch root and the net-section stress $\sigma^{\rm net}_{yy}$
\cite{Baratta70}. In PFC it is approximated by $K_t =\sigma^{\rm max}_{yy}/\sigma^{\rm net}_{yy}
\approx \varepsilon^{\rm max}_{yy}/\varepsilon^{\rm ref}_{yy}$ through an estimate from
linear elasticity \cite{PhysRevLett.96.225504}, where $\varepsilon^{\rm max}_{yy}$ and
$\varepsilon^{\rm ref}_{yy}$ are indicated in Fig.~\ref{fig:ff4}(h) showing the
cross-section profile of strain distribution in between the notches. Our IPFC simulation
gives $K_t = 1.991$, well agreeing with the value of $2.012$ calculated from the empirical
formula \cite{Baratta70} $K_t=(0.78+2.243\sqrt{L/R})[0.993+0.18(2L/W)-1.06(2L/W)^2+1.71(2L/W)^3]
(1-2L/W)$, where $W=21a_0$ and $L=R=3.5a_0$ in our setup (see Fig.~\ref{fig:ff1}).
Similar quantitative agreement has also been found in the previous MPFC study
\cite{PhysRevLett.96.225504}. As the sample is further stretched, cracks
are initiated at the tips of two notches with concentrated stresses and propagate
inside horizontally, as expected, causing the fracture of the nanoribbon. The
locations of the strain concentration propagate accordingly, as illustrated in
Figs.~\ref{fig:ff4}(e)--\ref{fig:ff4}(g).

\begin{figure*}
\includegraphics[width=0.9\textwidth]{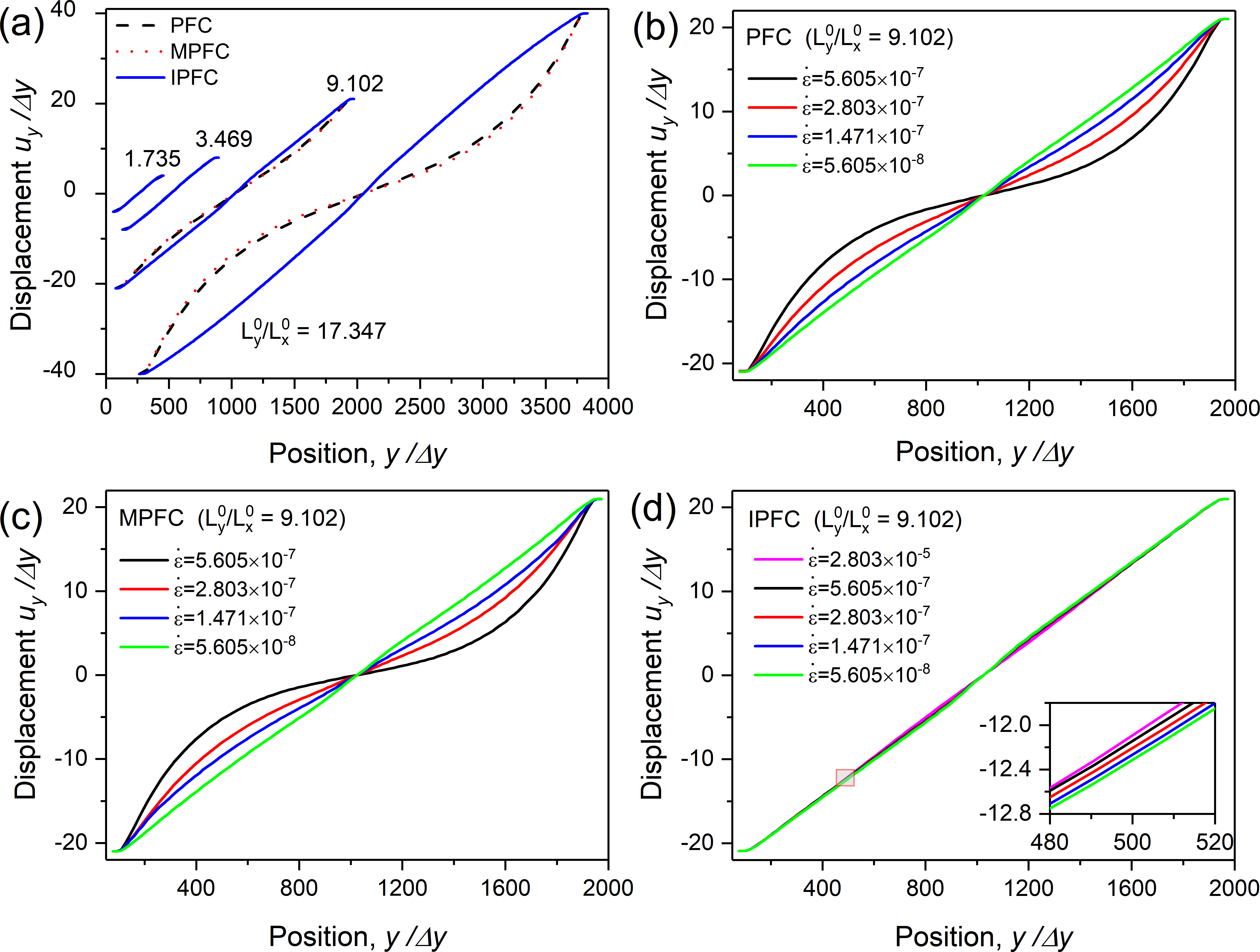}
\caption{The displacement $u_y$ along the tensile loading direction in the middle of the
  double notched nanoribbon simulated [e.g., the red vertical line in Fig.~\ref{fig:ff4}(a)],
  at strain $\varepsilon_e=2.35\%$.
  (a) The displacements in systems of different aspect ratios simulated by PFC,
  MPFC, and IPFC methods, for $L_y^0/L_x^0=1.735$ (with system size $256\Delta x 
  \times 512\Delta y$), $L_y^0/L_x^0=3.469$ (system size $256\Delta x \times 1024\Delta y$),
  $L_y^0/L_x^0=9.102$ (system size $256\Delta x \times 2048\Delta y$), and $L_y^0/L_x^0=17.347$
  (system size $256\Delta x \times 4096\Delta y$). In these systems $L_x^0 = 196 \Delta x$
  and $\dot{\varepsilon}=1.471 \times 10^{-7}$ are kept the same. In (b), (c), and (d) the
  double notched sample with $L_y^0/L_x^0=9.102$ is simulated under different applied strain
  rates, by using PFC, MPFC, and IPFC schemes, respectively.}
\label{fig:ff5}
\end{figure*}

It is important to note that while results from these three methods show similarities
at small system size and strain rate, they deviate more from each other at larger
length and/or faster rate of stretching. This can be seen clearly in Fig.~\ref{fig:ff5}
which shows the elastic response of the nanoribbon with different initial vertical length
$L_y^0$ (while $L_x^0$ is kept unchanged) or different applied strain rate $\dot{\varepsilon}$
under tensile deformation. For a system with total grid size $256 \times 512$ and
$L_y^0/L_x^0=340/196=1.735$ (as studied in Figs.~\ref{fig:ff3} and \ref{fig:ff4}),
at small strain the 1D profile of the displacement field $u_y$ exhibits the same linear
spatial dependence along the middle vertical line of the nanoribbon for PFC, MPFC, and
IPFC results, as shown in Fig.~\ref{fig:ff5}(a). A similar behavior applies to larger
size $256 \times 1024$ with $L_y^0/L_x^0=3.469$. However, when $L_y^0$ is further increased
(e.g., for grid size $256 \times 2048$ with $L_y^0/L_x^0=9.102$, and $256 \times 4096$
with $L_y^0/L_x^0=17.347$), at the same strain rate $\dot{\varepsilon}=1.471 \times 10^{-7}$
a deviation from the linear distribution of $u_y$ occurs for both the original PFC model
due to its slow, diffusive dynamics of relaxation, and the MPFC model as its elastic
interaction length has been exceeded at such a large length scale.

Similar effects can be found in terms of increasing strain rate $\dot{\varepsilon}$.
For both PFC and MPFC results given in Figs.~\ref{fig:ff5}(b) and \ref{fig:ff5}(c)
at $L_y^0/L_x^0=9.102$, it is shown that the fully mechanically relaxed state can always
be approached as long as the strain rate is sufficiently small (at the order of
$10^{-8}$ or less; i.e., if waiting for long enough time within each stretching step)
which, however, is computationally expensive particularly for large
systems. At faster but more realistic strain rates, a viscoelastic behavior of the
displacement field is obtained, instead of the elastic response, since the stressed
system dominated by diffusive processes (or with inadequate elastic relaxation mechanism)
needs long enough time to equilibrate elastically. Although in MPFC one can adjust
$\alpha$ and $\beta$ parameter values to reach larger length and time scales of elastic
interaction, too large $\alpha$ and/or too small $\beta$ would cause the difficulty of
numerical convergence (e.g., needing very small $\Delta t$) and reduce the computational
efficiency. In addition, it has been shown that there exists a lower limit of the
dissipative parameter $\beta$; below this limit no faster relaxation dynamics can
be gained \cite{HeinonenPRE16}.

In contrast, such restrictions of system size and strain rate are significantly released
for IPFC given its interpolation scheme, which instead can always lead to elastic
equilibrium of the system within a short time scale for all the system sizes [see
Fig.~\ref{fig:ff5}(a)] and strain rates [see Fig.~\ref{fig:ff5}(d)] examined in our tests.
Our calculations show that the IPFC scheme is at least an order of magnitude more
efficient than the original PFC and MPFC methods, giving a clear advantage of IPFC
for the study of mechanical deformation and response of solid systems.

\subsubsection{Deformation process during tensile test and fracture}

\begin{figure}
  \includegraphics[width=0.48\textwidth]{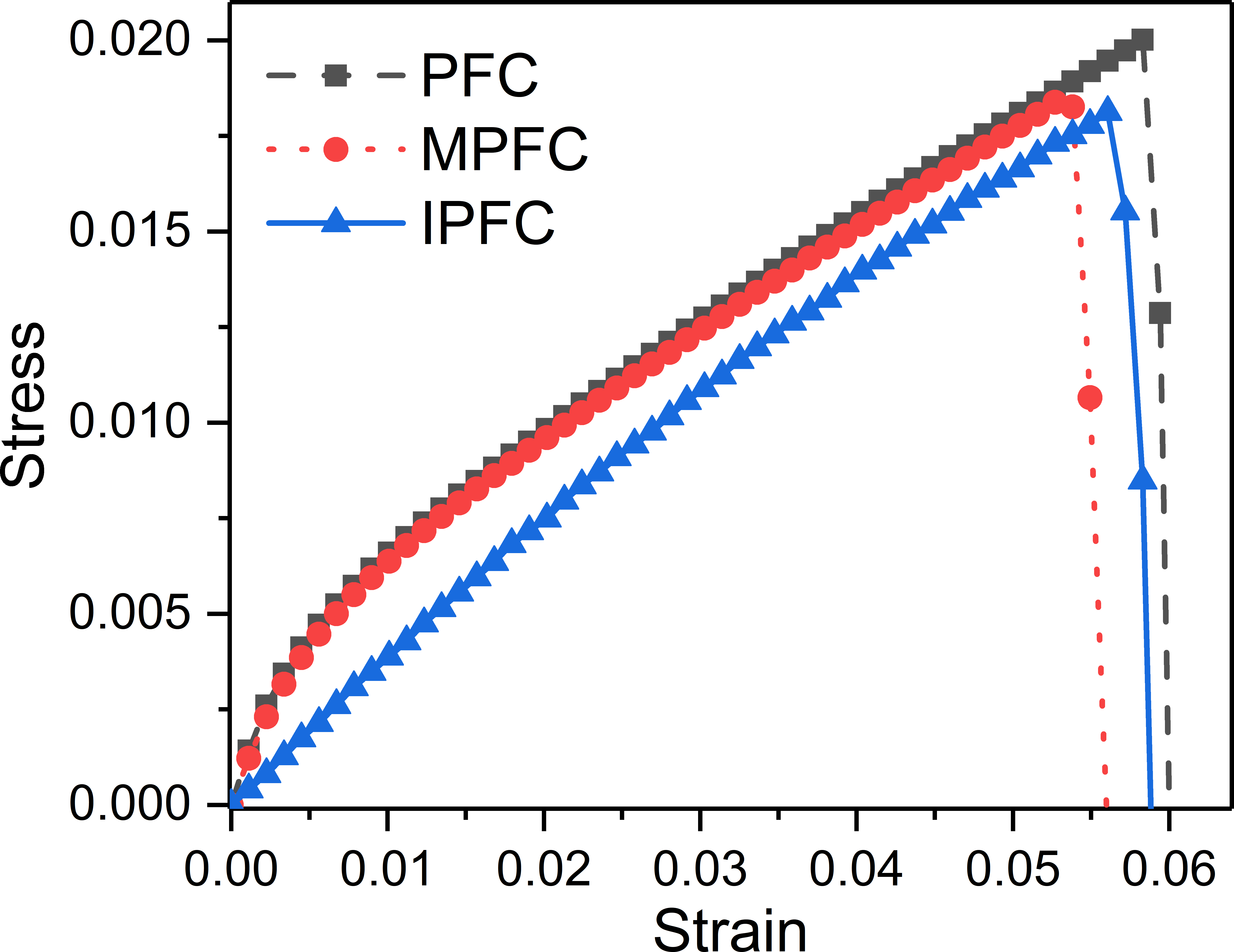}
\caption{Stress-strain curves obtained from three different methods, for $L_y^0/L_x^0 = 9.102$
  and $\dot{\varepsilon}=5.605 \times 10^{-7}$.}
\label{fig:ff6}
\end{figure}

To further investigate the detailed process of elastic deformation and fracture in the
uniaxially stressed nanoribbon, a system of long enough length along the pulling direction
with $L_y^0/L_x^0=1784/196=9.102$ (and grid size $256 \times 2048$) is simulated at the
same moderate strain rate $\dot{\varepsilon}=5.605 \times 10^{-7}$ using three PFC methods.
The calculated results of stress-strain relation are given in Fig.~\ref{fig:ff6}, showing
the discrepancies of mechanical property identified from original PFC, MPFC, and IPFC
in this large system. This is different from the small-system similarity presented in
Fig.~\ref{fig:ff3}. The discrepancies occur even in the early stage of small-strain
elastic regime (for which only IPFC yields the expected linear elastic behavior),
which can be attributed to different speeds of elastic relaxation in different methods
and the elastic vs viscoelastic behavior of the displacement field (see Fig.~\ref{fig:ff5}).
In addition, different values of fracture strain and strength are obtained through
three methods. 

\begin{figure}
\includegraphics[width=0.48\textwidth]{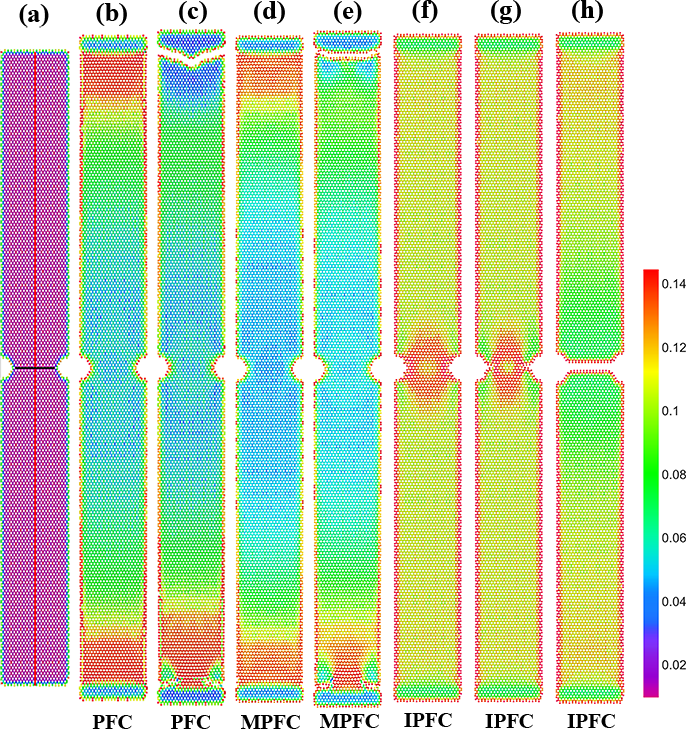}
\caption{Spatial distribution of strain $\varepsilon_{yy}$ over the double notched nanoribbon
  simulated in Fig.~\ref{fig:ff6}, at (a) $\varepsilon_e=0$, (b) $4.98\%$ and (c) $6.32\%$
  using PFC, (d) $5.02\%$ and (e) $5.89\%$ using MPFC, and (f) $5.45\%$, (g) $5.54\%$, and
  (h) $6.57\%$ using IPFC scheme.}
\label{fig:ff7}
\end{figure}

The corresponding results of color-coded spatial distribution of strain $\varepsilon_{yy}$
are illustrated in Fig.~\ref{fig:ff7}. In the case of original PFC model, the strain and stress
first concentrate in the vicinity of the traction regions and then propagate into the internal
of the sheet, but only partially due to the slow elastic propagation [see Fig.~\ref{fig:ff7}(b)].
This causes the over-concentrating of stress around the boundary with the loading region which
exceeds that of the notches, leading to the cracking near the edges of the traction regions
[Fig.~\ref{fig:ff7}(c)] instead of the notch roots. This abnormal fracture behavior also
occurs in the MPFC simulation given its limited range of elastic interaction, as shown in
Figs.~\ref{fig:ff7}(d) and \ref{fig:ff7}(e). In comparison, Figs.~\ref{fig:ff7}(f)--\ref{fig:ff7}(h)
show that as a result of fast elastic relaxation in IPFC, the strain distributes across the whole
sheet and concentrates around the notch tips, initializing the crack formation there and causing
the subsequent brittle cleavage fracture as expected in real materials.

\begin{figure*}
\includegraphics[width=0.9\textwidth]{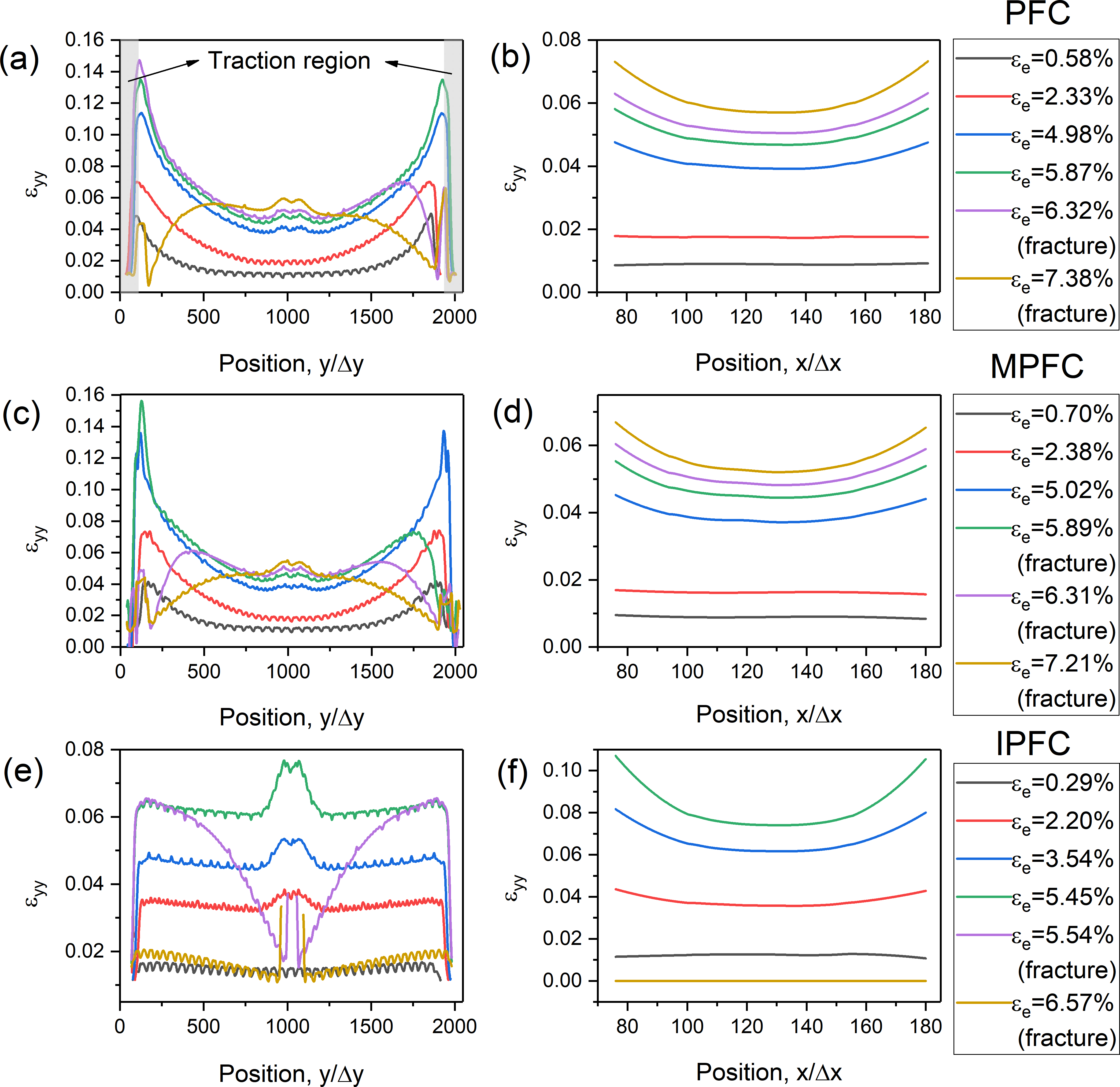}
\caption{Plots of $\epsilon_{yy}$ at different applied strain levels, either along the tensile
  loading direction in the middle of nanoribbon [the red vertical line in Fig.~\ref{fig:ff7}(a)]
  for (a), (c), (e), or perpendicular to the loading direction [the black central horizontal
  line in Fig.~\ref{fig:ff7}(a)] for (b), (d), (f).}
\label{fig:ff8}
\end{figure*}

These two different scenarios of fracture process can be further analyzed from the evolution
of 1D cross-section profiles of strain distribution plotted in Fig.~\ref{fig:ff8}, either
along the vertical pulling direction at $x=L_x/2$ [i.e., $\varepsilon_{yy}$ vs $y$ in
Fig.~\ref{fig:ff8} (a), (c), (e)], or perpendicular to the pulling at $y=L_y/2$ [i.e.,
$\varepsilon_{yy}$ vs $x$ in Fig.~\ref{fig:ff8} (b), (d), (f)]. Comparing the results
between three methods clearly shows how the discrepancy emerges among PFC, MPFC, and IPFC
schemes due to different strain propagation processes. For the $y$-direction profiles given
in Figs.~\ref{fig:ff8}(a) and \ref{fig:ff8}(c) for PFC and MPFC, although the strain inside
the traction regions is very small, consistent with our traction boundary condition setup,
the maxima of local strain always occur near the boundary between active and traction
regions before fracture (with $\varepsilon_e$ up to 5.87\% for PFC and 5.05\% for MPFC
in the figures), which can even reach a value close to 14\% for $\varepsilon_{yy}$.
This contradicts the usual expectation of stress concentration around the notch region
located in the middle of the nanoribbon (which instead shows close-to-minimum strain).
It indicates an outcome of inadequate mechanical relaxation in these two methods which
requires a much longer time for the propagation of the imposed strain and stress into the
internal of this large system. Such large strains concentrated at the active-traction
boundaries eventually lead to the cracking and fracture there, showing as a rapid decrease
of strain values at the crack locations [see strain profiles of $\varepsilon_e=6.32\%$
(purple) and 7.38\% (yellow) in Fig.~\ref{fig:ff8}(a) for PFC, and 5.89\% (green),
6.31\% (purple), and 7.21\% (yellow) in Fig.~\ref{fig:ff8}(c) for MPFC].

A qualitatively different behavior is observed in the results of IPFC. As shown in
Fig.~\ref{fig:ff8}(e), within each stretching step the stress applied at the boundaries
has well propagated into the bulk, and before the occurrence of fracture the strain
distribution peaks around the middle ($y=L_y/2$) where the notch section is located,
demonstrating the efficiency of IPFC scheme in terms of fast elastic relaxation when
subjected to mechanical deformation. Within the notch section the strain is concentrated
at the roots as expected, i.e., at the two ends of the $x$-direction profiles given in
Fig.~\ref{fig:ff8}(f). The corresponding stress-concentration factor $K_t$, which can be
approximated as the ratio between maximum and minimum values of $\varepsilon_{yy}$ along
the $\varepsilon_{yy}$ vs $x$ profile, is larger than that of PFC and MPFC plots shown in
Figs.~\ref{fig:ff8}(b) and \ref{fig:ff8}(d). Cracks are then initialized at the notch tips
and the brittle fracture occurs, which is associated with the steep drop of strain in the
middle notch region of Fig.~\ref{fig:ff8}(e) and the zero strain value across the central
notch section in Fig.~\ref{fig:ff8}(f) when $\varepsilon_e=5.54\%$ (purple) and 6.57\%
(yellow).

\subsection{Temperature and chirality dependence of mechanical property}

Given the efficiency of the IPFC scheme as demonstrated above, we use it to systematically
examine the mechanical property of the double notched nanoribbon with honeycomb lattice
structure, particularly effects of system temperature and chirality. Noting that the PFC
free energy functional Eq.~(\ref{1101}) with honeycomb symmetry has been used as an
effective approach for the study of 2D graphene monolayers \cite{Elder16,Smirman17},
our calculations are expected to reveal some important mechanical properties of graphene.

\begin{figure}
  \includegraphics[width=0.45\textwidth]{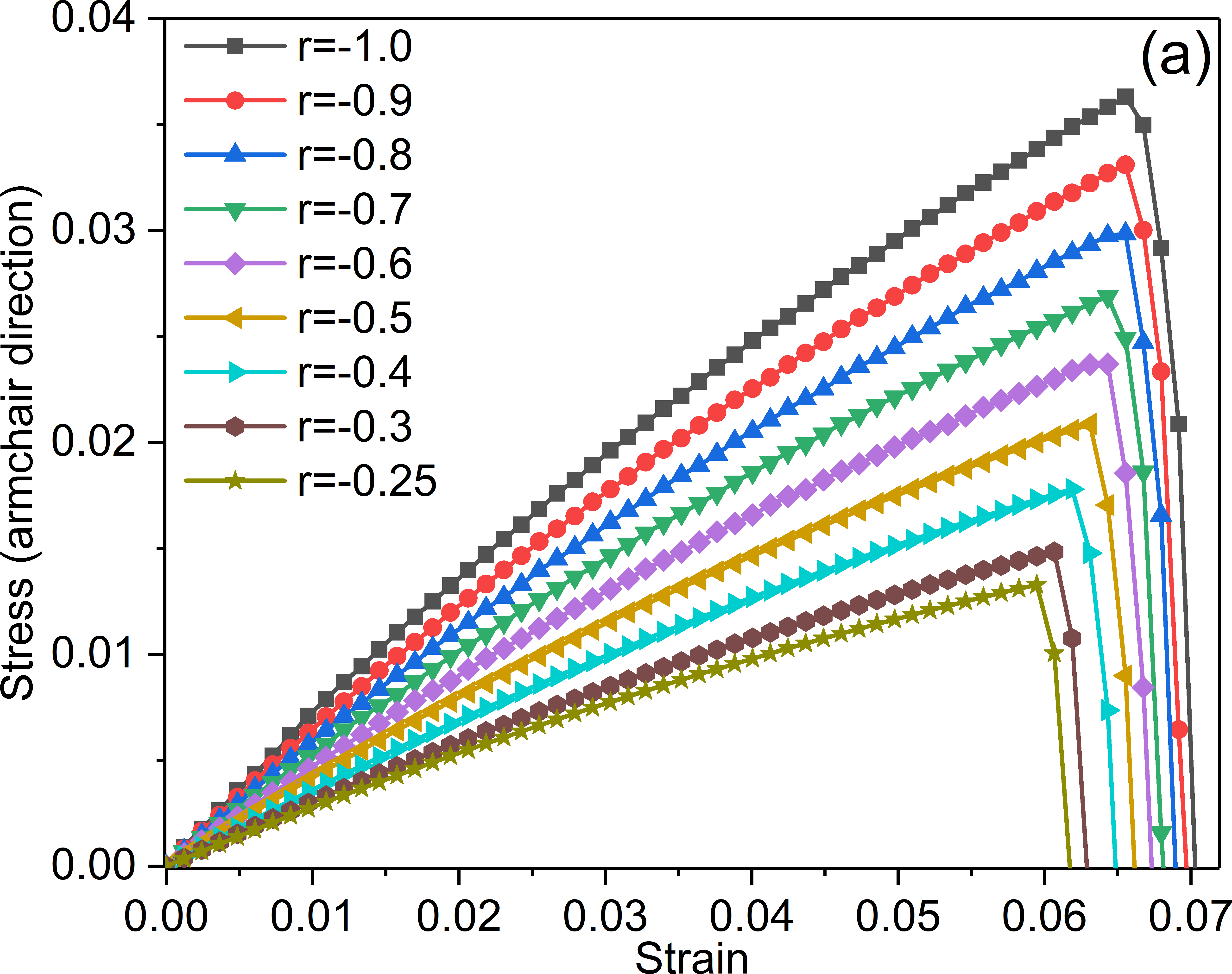}\\
  \includegraphics[width=0.45\textwidth]{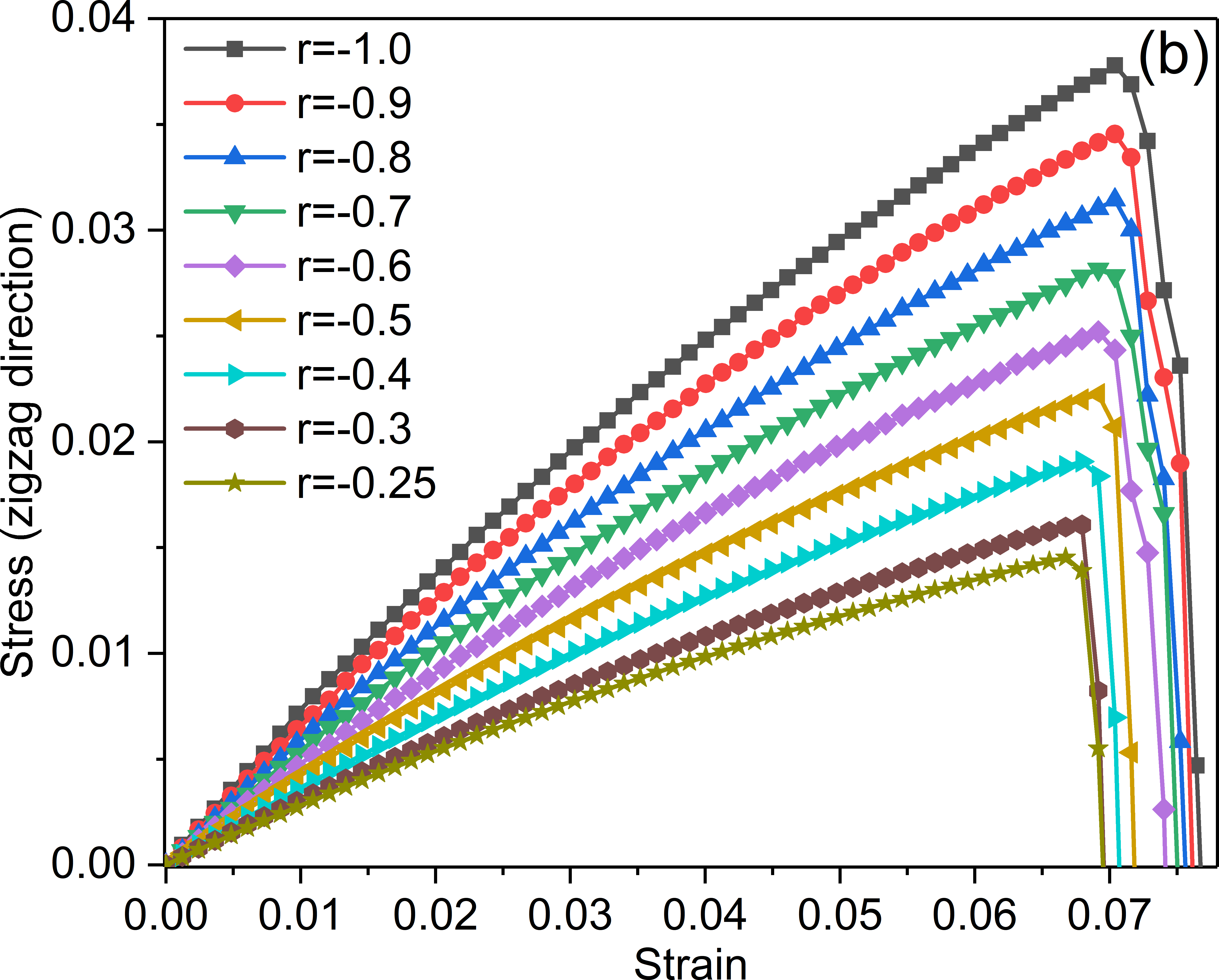}
  \caption{Temperature-dependent stress-strain relation at different values of temperature
    parameter $r$, for the double notched nanoribbon system under uniaxial tension along
    the armchair (a) and zigzag (b) directions. In each simulation the total system size is
    $256 \Delta x \times 2048 \Delta y$, with $L_y^0/L_x^0=8.408$, and the applied strain
    rate is $\dot{\varepsilon}=6.068 \times 10^{-7}$.}
  \label{fig:ff9}
\end{figure}

All the results given above are for the uniaxial tension along the armchair direction
of honeycomb structure (see Fig.~\ref{fig:ff1}) at a fixed temperature parameter value
$r=-0.5$. More general results of stress-strain relation are presented in Fig.~\ref{fig:ff9},
as obtained from IPFC simulations for stretching along both armchair and zigzag directions
at various temperatures. Note that the temperature parameter $r$ is related to the
distance from the melting point, and greater value of $r$ corresponds to higher temperature.
In our simulations its largest value (i.e., $r=-0.25$) is chosen such that the double
notched nanoribbon can still maintain its faceted surface configuration during the mechanical
deformation. Results across different temperatures (i.e., different $r$ values) give
qualitatively similar behaviors of mechanical response and brittle fracture, as shown in
Figs.~\ref{fig:ff9}(a) and \ref{fig:ff9}(b) for uniaxial tensile tests along the armchair
and zigzag directions, respectively, although the quantitative outcomes are different in
both linear and nonlinear elastic and fracture regimes.

\begin{figure}
  \includegraphics[width=0.45\textwidth]{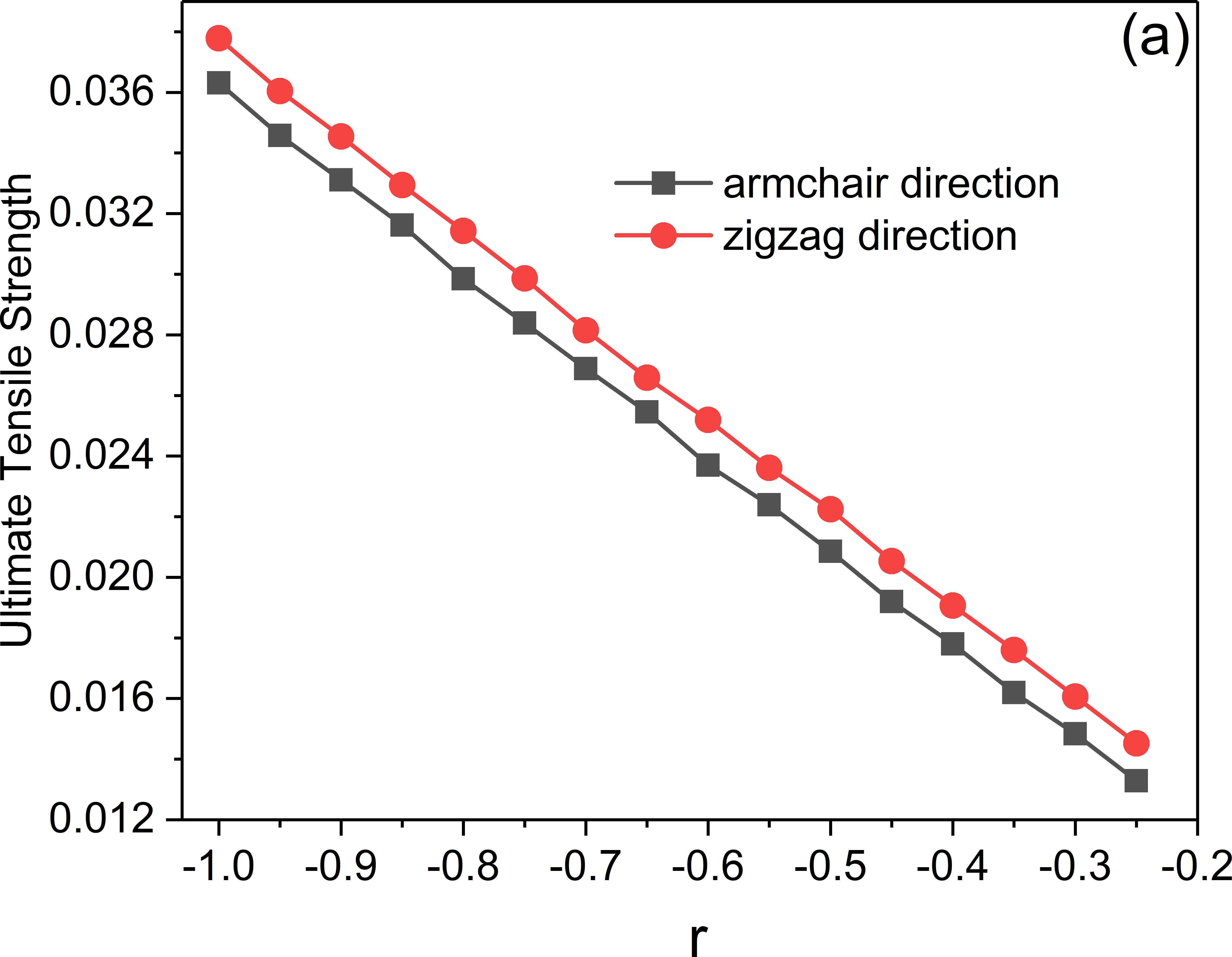}\\
  \includegraphics[width=0.45\textwidth]{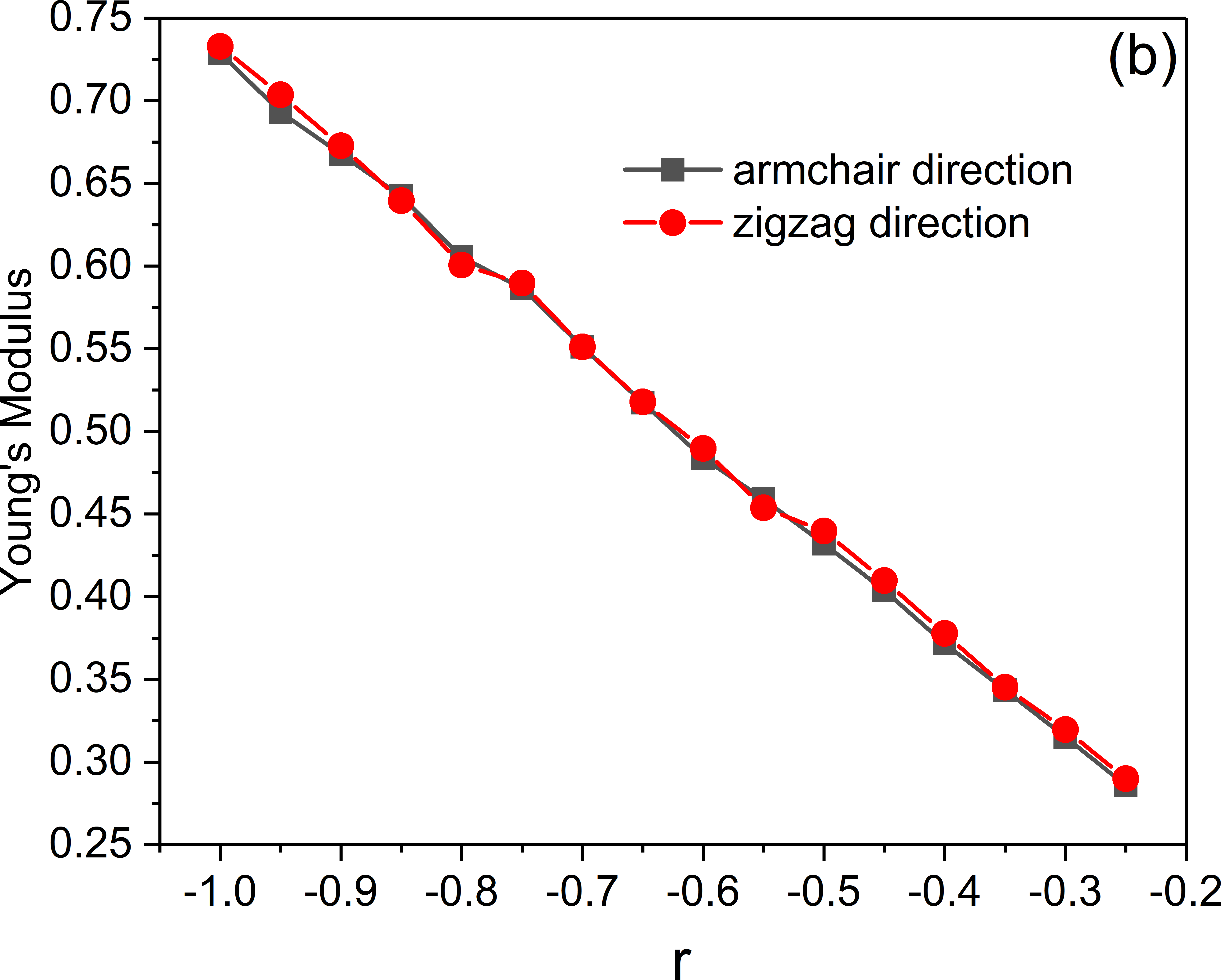}
  \caption{(a) Ultimate tensile strength as a function of temperature parameter $r$.
    (b) Young's modulus as a function of $r$. The results are determined from the
    stress-strain relation given in Fig.~\ref{fig:ff9}.}
  \label{fig:ff10}
\end{figure}

The corresponding values of temperature-varying ultimate tensile strength and Young's
modulus are presented in Fig.~\ref{fig:ff10}. The ultimate strength is identified
as the value of maximum stress before fracture, while the Young's modulus,
$Y = (\partial \sigma_e / \partial \varepsilon_e)_{\varepsilon_e \rightarrow 0}$,
is calculated from the linear elasticity regime of the stress-strain curve at small
strains (i.e., the slope of the stress-strain line for $\varepsilon_e< 0.5\%$).
Fig.~\ref{fig:ff10}(a) shows the decrease of the ultimate strength with increasing
temperature (i.e., increasing $r$), which is consistent with the previous MD results for
bulk pristine graphene simulated across a broad range of temperatures \cite{ZhaoJAP10}.
The similar temperature weakening behavior is also obtained for Young's modulus [see
Fig.~\ref{fig:ff10}(b)]. This can be attributed to the weakened
interparticle interaction strength and hence the softening of crystal at higher temperature,
as found in recent MD \cite{ZhaoJAP10} and MC \cite{ZakharchenkoPRL09} calculations
which showed the decrease of Young's modulus of graphene at high enough temperatures.

Figure \ref{fig:ff10}(a) also indicates slightly higher ultimate strengths in the zigzag
direction as compared to the armchair direction across the whole temperature range simulated,
a result of chirality effect that is consistent with the findings in both MD \cite{ZhaoNanoLett09}
and \textit{ab initio} DFT \cite{liu07} calculations of pristine graphene. A different
behavior is given in Fig.~\ref{fig:ff10}(b) for Young's modulus, showing very close values
along the armchair and zigzag directions. It is different from the MD result of graphene
nanoribbons \cite{ZhaoNanoLett09} which yielded higher Young's modulus along the zigzag
direction than the armchair one. This discrepancy can be understood from the fact that
the out-of-plane deformations have been incorporated in the MD simulation of free-standing
graphene sheets, but are neglected in our PFC simulations which are restricted to purely
2D planar structures. Under large degree of lateral stretching (particularly close to
fracture with high tensile strain) the stretched sheet would be flattened and any initial
out-of-plane variations would then play a negligible role, resulting in the consistency between
3D MD and 2D IPFC results for the chirality dependence of ultimate strength. However, at very
small strains used to calculate Young's modulus, these variations would influence the evaluation
of elastic properties, as seen in the discrepancy described above. This is also consistent
with the previous atomistic study of graphene \cite{ZhaoNanoLett09}, which showed that the
purely 2D tight-binding calculations yielded close magnitudes of Young's modulus for
armchair and zigzag directions, similar to our 2D IPFC results but different from those
of 3D MD calculations involving out-of-plane effects.

\section{Conclusions}

We have developed a computational method to effectively simulate the processes of
mechanical relaxation and fracture in PFC, by applying an interpolation scheme on
the PFC density field based on an imposed condition for the displacement field to
ensure local elastic equilibration and fast global mechanical relaxation. This IPFC
method is applied to the simulation of a sample 2D system with honeycomb lattice
structure under uniaxial tension. The results are compared to those obtained by
original PFC and MPFC simulations and to previous MD, MC, and \textit{ab initio}
DFT calculations of graphene. The IPFC-calculated stress-strain relation, ultimate
strength of brittle fracture and Young's modulus are qualitatively consistent with
those of atomistic simulations \cite{ZhaoNanoLett09,ZhaoJAP10,liu07,ZakharchenkoPRL09},
in terms of effects of system temperature (for both ultimate strength and Young's
modulus) and chirality (armchair vs zigzag direction, for ultimate strength and also
Young's modulus of 2D calculations). The outcomes demonstrate a much more efficient
process of dynamic relaxation and mechanical equilibration for IPFC scheme, in
comparison to the original PFC and MPFC methods. Although the use of this interpolation
scheme is still subjected to the scales of conventional PFC-type simulations, its
advantage applies particularly to the scenarios with large enough system sizes and/or
high enough strain rates, where the original PFC and MPFC models would generate
qualitatively incorrect results of mechanical response due to their inadequate mechanisms
for strain propagation and mechanical relaxation, including the stress-strain relation,
spatial distribution of strain or stress, spatial dependence of the displacement
field even in the linear elastic regime, and the fracture behavior.

It is noted that although this IPFC numerical scheme bears some similarity of treatment
as compared to a recently developed mechanically-equilibrated amplitude model \cite{Heinonen14}
in terms of imposing extra constraints on PFC dynamics, the detailed setup of constraints
is very different and the algorithm developed here is mainly for the study of mechanical
deformation and fracture. In the amplitude model \cite{Heinonen14} the equilibrium
condition was imposed only on the phases of amplitudes, while in our IPFC scheme the
interpolation is applied to the whole density field (equivalent to both magnitudes
and phases). Importantly, our method is applied to the full model of PFC instead of the
amplitude expansion at slow scales, so that the key features of original PFC model are
maintained, particularly the coupling between microscopic and mesoscopic length scales
\cite{Huang13} and the resulting effects of lattice pinning and Peierls barriers for
dislocation defects that are missing in the amplitude expansion but important for
examining the mechanical response of materials. Even when directly applying this IPFC
scheme to the amplitude equations (which would be straightforward), the interpolation
would be imposed on both the average density $\phi_0$ and the full complex amplitudes
$A_{nm}$ but not only on their phases. Note also that the interpolation algorithm
[Eqs.~(\ref{2094}) and (\ref{2095})] used here for uniaxial tensile test can be
extended straightforwardly to the study of compressive test as well as biaxial
tension/compression of materials. In addition, this IPFC scheme is independent
of the model free energy adopted and thus can be directly used for any other PFC
models with more complex free energy functionals (see, e.g.,
Refs. \cite{PhysRevLett.105.045702,Mkhonta13,Wang18r,PhysRevLett.118.255501}) for the
modeling of a wider variety of material systems with different crystalline symmetries
and microstructures.

\begin{acknowledgments}
  Z.-F.H. acknowledges support from the National Science Foundation under Grant
  No. DMR-1609625. J.W. and Z.W. acknowledge support from the National Natural Science
  foundation of China (Grants No. 51571165, 51371151), and the Center for High Performance
  Computing of Northwestern Polytechnical University, China for computer time and facilities.
\end{acknowledgments}

\appendix*
\section{Numerical method for solving MPFC equation}

In the following we outline the derivation of a pseudospectral algorithm for solving the
MPFC dynamic equation (\ref{1121}). We follow the method of Ref. \cite{RevModPhys.15.1}
for the formal solution of an ordinary differential equation (now in Fourier space) with
second-order time derivative, as adopted in Ref. \cite{PhysRevB.87.024110}. Compared
to the explicit numerical scheme derived in Ref. \cite{PhysRevB.87.024110} for MPFC,
here we (i) express the equations of algorithm in terms of real quantities (using
hyperbolic functions), and (ii) use an implicit treatment for nonlinear terms and
the predictor-corrector method \cite{Cross94}.

In Fourier space the MPFC equation (\ref{1121}) is written as
\begin{equation}\label{22}
  \frac{\partial^2 \hat{\phi}_{\vec{q}}}{\partial t^2}
  + \beta \frac{\partial \hat{\phi}_{\vec{q}}}{\partial t}
  = \sigma_q \hat{\phi}_{\vec{q}} + \hat{N}_{\vec{q}},
\end{equation}
where $\hat{\phi}_{\vec{q}}$ is the Fourier component of the density field $\phi$,
$\sigma_q = -\alpha^2 q^2 [r+(1-q^2)^2]$ with wave number $q$, and $\hat{N}_{\vec{q}}$
is the Fourier transform of the nonlinear terms $\alpha^2\nabla^2(\tau\phi^2+\phi^3)$.
The general solution of Eq.~(\ref{22}) is of the form
\begin{equation}\label{55}
\hat{\phi}_{\vec{q}}=a_1(t)e^{\frac{1}{2}(-\beta+\beta_1)t}+a_2(t)e^{\frac{1}{2}(-\beta-\beta_1)t},
\end{equation}
where $\beta_1=\sqrt{\beta^2+4\sigma_q}$, and $a_1$ and $a_2$ satisfy
the condition \cite{RevModPhys.15.1} $(da_1/dt)\exp[\frac{1}{2}(-\beta+\beta_1)t]
+(da_2/dt)\exp[\frac{1}{2}(-\beta-\beta_1)t]=0$, such that
\begin{equation}\label{66}
\frac{da_1}{dt}e^{\beta_1t}+\frac{da_2}{dt}=0.
\end{equation}

Substituting Eq.~(\ref{55}) into Eq.~(\ref{22}) and combining with Eq.~(\ref{66}),
we get
\begin{eqnarray}
&& \frac{da_1}{dt}=\frac{1}{\beta_1} \hat{N}_{\vec{q}}(t) e^{\frac{1}{2}(\beta-\beta_1)t},\\
&& \frac{da_2}{dt}=-\frac{1}{\beta_1} \hat{N}_{\vec{q}}(t) e^{\frac{1}{2}(\beta+\beta_1)t}.
\end{eqnarray}
Integrating them from $t$ to $t+\Delta t$ gives
\begin{eqnarray}
  a_1(t+\Delta t) &=& \frac{1}{\beta_1} \int_{t}^{t+\Delta t} dt^\prime
  e^{\frac{1}{2}(\beta-\beta_1)t^\prime} \hat{N}_{\vec{q}}(t^\prime) + a_1(t),~~\quad \label{88}\\
  a_2(t+\Delta t) &=& -\frac{1}{\beta_1} \int_{t}^{t+\Delta t} dt^\prime
  e^{\frac{1}{2}(\beta+\beta_1)t^\prime} \hat{N}_{\vec{q}}(t^\prime) + a_2(t).~~\quad \label{99}
\end{eqnarray}
For a given time $t$, $a_1(t)$ and $a_2(t)$ can be expressed in terms of $\hat{\phi}_{\vec{q}}$
and its time derivative $\hat{u}_{\vec{q}}$, which is determined from Eqs.~(\ref{55}) and
(\ref{66}) to be
\begin{eqnarray}
  \hat{u}_{\vec{q}}(t)  = \frac{\partial \hat{\phi}_{\vec{q}}}{\partial t}
  &=& \frac{1}{2}(-\beta+\beta_{1})a_1(t)e^{\frac{1}{2}(-\beta+\beta_{1})t} \nonumber\\
  &-& \frac{1}{2}(\beta+\beta_{1})a_2(t)e^{-\frac{1}{2}(\beta+\beta_{1})t}.
\label{77}
\end{eqnarray}
Combining Eqs.~(\ref{55}) and (\ref{77}) leads to
\begin{equation}
  a_1(t) = \frac{1}{\beta_1} e^{\frac{1}{2}(\beta-\beta_1)t} \left [ \hat{u}_{\vec{q}}(t)
    + \frac{1}{2}(\beta+\beta_1) \hat{\phi}_{\vec{q}}(t) \right ], \label{eq:a1}
\end{equation}
and
\begin{equation}
  a_2(t) = -\frac{1}{\beta_1} e^{\frac{1}{2}(\beta+\beta_1)t} \left [ \hat{u}_{\vec{q}}(t)
    + \frac{1}{2}(\beta-\beta_1) \hat{\phi}_{\vec{q}}(t) \right ]. \label{eq:a2}
\end{equation}

Evaluating Eqs.~(\ref{55}) and (\ref{77}) at $t+\Delta t$ and using Eqs.~(\ref{88}) and
(\ref{99}) for $a_1(t+\Delta t)$ and $a_2(t+\Delta t)$, we have
\begin{eqnarray}
  \hat{\phi}_{\vec{q}}(t+\Delta t) &=&
  \frac{2}{\beta_1} \int_{t}^{t+\Delta t} dt^\prime e^{-\frac{1}{2}\beta(t+\Delta t-t^\prime)} \nonumber\\
  && \times \sinh \left [ \frac{\beta_1}{2}(t+\Delta t-t^\prime) \right ]
  \hat{N}_{\vec{q}}(t^\prime) \nonumber\\
  &+& a_1(t) e^{-\frac{1}{2}(\beta-\beta_1)(t+\Delta t)} \nonumber\\
  &+& a_2(t) e^{-\frac{1}{2}(\beta+\beta_1)(t+\Delta t)}, \label{eq:phiq}
\end{eqnarray}
and
\begin{eqnarray}
  \hat{u}_{\vec{q}}(t+\Delta t) &=& \frac{1}{\beta_1} \int_{t}^{t+\Delta t} dt^\prime
  e^{-\frac{1}{2}\beta(t+\Delta t-t^\prime)} \nonumber\\
  && \times \left \{ -\beta \sinh \left [ \frac{\beta_1}{2} (t+\Delta t-t^\prime) \right ]
  \right. \nonumber\\
  && \left. \quad +\beta_1 \cosh \left [ \frac{\beta_1}{2} (t+\Delta t-t^\prime) \right ] \right \}
  \hat{N}_{\vec{q}}(t^\prime) \nonumber\\
  &+& \frac{1}{2}(-\beta+\beta_1) a_1(t) e^{\frac{1}{2}(-\beta+\beta_1)(t+\Delta t)} \nonumber\\
  &-& \frac{1}{2}(\beta+\beta_1) a_2(t) e^{-\frac{1}{2}(\beta+\beta_1)(t+\Delta t)}. \label{eq:uq}
\end{eqnarray}
For an implicit scheme, the nonlinear term $\hat{N}_{\vec{q}}(t^\prime)$ is expanded to the first
order as \cite{Cross94}
\begin{equation}
  \hat{N}_{\vec{q}}(t^\prime)=\hat{N}_{\vec{q}}(t)
  +\frac{\hat{N}_{\vec{q}}(t+\Delta t)-\hat{N}_{\vec{q}}(t)}{\Delta t}(t^\prime-t). \label{eq:Nq}
\end{equation}
Substituting Eq. (\ref{eq:Nq}) into Eqs. (\ref{eq:phiq}) and (\ref{eq:uq}), integrating over
$t^\prime$, and making use of Eqs. (\ref{eq:a1}) and (\ref{eq:a2}), we obtain the following
numerical algorithm for solving the MPFC dynamic equation.

\subsection{When $\sigma_q \neq 0$}

If $\beta_1$ is real and positive (i.e., $\beta^2+4\sigma_q>0$), we have
\begin{widetext}
\begin{eqnarray}
  \hat{\phi}_{\vec{q}}(t+\Delta t) &=& \hat{\phi}_{\vec{q}}(t) e^{-\frac{1}{2}\beta\Delta t}
  \left [ \frac{\beta}{\beta_1} \sinh \left (\frac{1}{2}\beta_1\Delta t \right )
    +\cosh \left ( \frac{1}{2}\beta_1\Delta t \right ) \right ]
  +\frac{2}{\beta_1} \hat{u}_{\vec{q}} e^{-\frac{1}{2}\beta\Delta t}
  \sinh \left ( \frac{1}{2}\beta_1\Delta t \right ) \nonumber\\
  && + \frac{\hat{N}_{\vec{q}}(t)}{\sigma_q} \left \{ e^{-\frac{1}{2}\beta\Delta t}
  \left [ \frac{\beta}{\beta_1} \sinh \left ( \frac{1}{2}\beta_1\Delta t \right )
    + \cosh \left ( \frac{1}{2}\beta_1\Delta t \right ) \right ] -1 \right \} \nonumber\\
  && + \frac{\hat{N}_{\vec{q}}(t+\Delta t)-\hat{N}_{\vec{q}}(t)}{\sigma_q^2 \Delta t}
  \left \{ e^{-\frac{1}{2}\beta\Delta t} \left [ \frac{\beta^2+\beta_1^2}{2\beta_1}
    \sinh \left ( \frac{1}{2}\beta_1\Delta t \right ) + \beta\cosh \left (
    \frac{1}{2}\beta_1\Delta t \right ) \right ] -\beta -\sigma_q\Delta t \right \},
  \label{eq:phi}
\end{eqnarray}
and
\begin{eqnarray}
  \hat{u}_{\vec{q}}(t+\Delta t) &=& \frac{2\sigma_q}{\beta_1} \hat{\phi}_{\vec{q}}(t)
  e^{-\frac{1}{2}\beta\Delta t} \sinh \left ( \frac{1}{2}\beta_1\Delta t \right )
  + \hat{u}_{\vec{q}}(t) e^{-\frac{1}{2}\beta\Delta t} \left [-\frac{\beta}{\beta_1}
    \sinh \left ( \frac{1}{2}\beta_1\Delta t \right )
    +\cosh \left ( \frac{1}{2}\beta_1\Delta t \right ) \right ] \nonumber\\
  && + \frac{2\hat{N}_{\vec{q}}}{\beta_1} e^{-\frac{1}{2}\beta\Delta t}
  \sinh \left ( \frac{1}{2}\beta_1\Delta t \right ) \nonumber\\
  && + \frac{\hat{N}_{\vec{q}}(t+\Delta t)-\hat{N}_{\vec{q}}(t)}{\sigma_q \Delta t}
  \left \{ e^{-\frac{1}{2}\beta\Delta t} \left [ \frac{\beta}{\beta_1}
    \sinh \left ( \frac{1}{2}\beta_1\Delta t \right ) +
    \cosh \left ( \frac{1}{2}\beta_1\Delta t \right ) \right ] -1 \right\}.
  \label{eq:u}
\end{eqnarray}
\end{widetext}

If $\beta_1 = i\alpha_1$ is imaginary, i.e., $\beta^2+4\sigma_q<0$ and
$\alpha_1=\sqrt{-(\beta^2+4\sigma_q)}$, we only need to replace the terms
$\frac{1}{\beta_1}\sinh(\frac{1}{2}\beta_1\Delta t)$ and $\cosh(\frac{1}{2}\beta_1\Delta t)$
in the above equations via
\begin{eqnarray}
  && \frac{1}{\beta_1}\sinh \left ( \frac{1}{2}\beta_1\Delta t \right )
  =\frac{1}{\alpha_1} \sin \left (\frac{1}{2}\alpha_1\Delta t \right ), \nonumber\\
  && \cosh \left ( \frac{1}{2}\beta_1\Delta t \right )
  =\cos \left ( \frac{1}{2}\alpha_1\Delta t \right ).
\end{eqnarray}
In the case of $\beta_1=0$ (with $\beta^2+4\sigma_q=0$), those terms become
\begin{equation}
  \frac{1}{\beta_1} \sinh \left ( \frac{1}{2}\beta_1\Delta t \right )
  \rightarrow \frac{1}{2}\Delta t, \quad
  \cosh \left ( \frac{1}{2}\beta_1\Delta t \right ) = 1.
\end{equation}

\subsection{When $\sigma_q = 0$}

In this case we have $\beta_1=\beta>0$, and thus Eqs.~(\ref{eq:phi}) and (\ref{eq:u})
are replaced by
\begin{eqnarray}
  && \hat{\phi}_{\vec{q}}(t+\Delta t) = \hat{\phi}_{\vec{q}}(t)
  +\frac{1-e^{-\beta\Delta t}}{\beta} \hat{u}_{\vec{q}}(t) \nonumber\\
  && +\hat{N}_{\vec{q}}(t) \left ( \frac{\Delta t}{\beta}
  +\frac{e^{-\beta\Delta t}-1}{\beta^2} \right ) \nonumber\\
  && + \left [ \hat{N}_{\vec{q}}(t+\Delta t)-\hat{N}_{\vec{q}}(t) \right ]
  \left ( \frac{\Delta t}{2\beta}-\frac{e^{-\beta\Delta t}-1+\beta\Delta t}{\beta^3\Delta t}
  \right ), \nonumber\\ \label{eq:phi_0}
\end{eqnarray}
\begin{eqnarray}
  && \hat{u}_{\vec{q}}(t+\Delta t) = \hat{u}_{\vec{q}}(t)e^{-\beta\Delta t}
  -\hat{N}_{\vec{q}}(t)\frac{e^{-\beta\Delta t}-1}{\beta} \nonumber\\
  && \qquad + \left [ \hat{N}_{\vec{q}}(t+\Delta t)-\hat{N}_{\vec{q}}(t) \right ]
  \frac{e^{-\beta\Delta t}-1+\beta\Delta t}{\beta^2\Delta t}. \label{eq:u_0}
\end{eqnarray}

To implement the implicit numerical scheme described above, we use the predictor-corrector
method as in Ref.~\cite{Cross94}. For the predictor step $\hat{N}_{\vec{q}}(t+\Delta t)
-\hat{N}_{\vec{q}}(t)$ is assumed to be zero in Eqs.~(\ref{eq:phi}) and (\ref{eq:u}) or
Eqs.~(\ref{eq:phi_0}) and (\ref{eq:u_0}), yielding the predictor or guess values of
$\hat{\phi}_{\vec{q}}(t+\Delta t)$ and $\hat{u}_{\vec{q}}(t+\Delta t)$, while at the corrector
step these guess values are used to evaluate $\hat{N}_{\vec{q}}(t+\Delta t)$ and thus the
updated values of $\hat{\phi}_{\vec{q}}(t+\Delta t)$ and $\hat{u}_{\vec{q}}(t+\Delta t)$.
At each time $t+\Delta t$, in principle the corrector step would include multiple iterations
until the result reaches the desired numerical accuracy, although in our simulations we
conduct only one predictor-corrector iteration which is sufficient for numerical convergence
with enough accuracy of the outcomes.

\bibliography{mybibfile}

\end{document}